\documentclass[12pt,tightenlines,eqsecnum,floats,showpacs,nofootinbib,amsmath,amssymb,aps]{revtex4}

\usepackage{amsmath,amssymb,amsfonts,amsthm,amscd}
\usepackage{graphicx}
\usepackage{enumerate}
\usepackage{colordvi}
\usepackage{hyperref}
\usepackage[hang, flushmargin]{footmisc}
\usepackage{natbib}
\usepackage{enumerate}
\usepackage{color}
\def\be{\begin{equation}}
\def\ee{\end{equation}}
\def\ba{\begin{eqnarray}}
\def\ea{\end{eqnarray}}

\def\Lie{\mathbf{L}}
\def\L{\mathcal{L}}
\def\S{\mathcal{S}}
\def\f{\frac}

\def\H{\mathcal{H}}

\begin{document}

\title{Minimal Coupling and Attractors}
\author{David Sloan$^1$}
\email{djs228@hermes.cam.ac.uk}
\affiliation{$^{1}$ DAMTP, Centre for Mathematical Sciences, Wilberforce Rd.,
Cambridge
University, Cambridge CB3 0WA, UK}

\begin{abstract}

The effects of minimally coupling a gravity to matter on a flat Robertson-Walker geometry are explored. Particular attention is paid to the evolution of the symplectic structure and the Liouville measure it defines. We show that the rescaling freedom introduced by choice of fiducial cell leads to a symmetry between dynamical trajectories, which together with the Liouville measure provides a natural volume weighting explanation for the generic existence of attractors. 

\end{abstract}

\maketitle

\section{Introduction}

Minimal coupling represents the simplest manner of adding matter to a gravitational theory. This is achieved, in effect, by taking a matter action and replacing ordinary derivatives with their covariant counterparts, raising and lowering indices with the metric and performing integrals over space-time using a volume form determined by the metric. The resulting theory will obey the equivalence principle and principle of general covariance if the gravitational theory does. When restricted to the flat ($k=0$) Robertson-Walker geometries, the homogeneous and isotropic models on which much of cosmology is founded, further simplifications occur. Since space is homogeneous only temporal derivatives play a role in our theory, and thus all the covariant derivatives remaining are simply derivatives with respect to some choice of time coordiante. Furthermore, any choice of of scale factor (or equivalently volume) is determined only up to a choice of overall scale, since physical parameters must be measured with respect to some fiducial cell whose size should play no role in dynamics. It can be argued that if our spatial manifold is compact, such as a three-torus say, then this structure could provide a length scale against which futher measurement can be based. However from the perspective of an observer who only has access to homogeneous local field configurations there is no physical observation of the fields from which this length scale can be deduced. This freedom to rescale leads to a symmetry on the space of solutions. This has been discussed in the context of inflation in Loop Quantum Cosmology \cite{Measure,Measure2} and its result in explaining inflationary attractors in \cite{Corichi:2013kua}.

The existence of attractors extends beyond the relatively simple context of single field inflation in general relativity. Here we will show how attractors arise in generic theories of gravity minimally coupled to matter as a result of the rescaling invariance. The restriction to minimally coupled systems, particularly those which obey the weak energy condition, will be made to allow for certain technical constructions, such as the monotonicity of a field variable (the conjugate momentum to volume) which will be used in evaluating the area of phase space occupied by sets of dynamical trajectories. As will be shown, this is not strictly required, but does simplify matters considerably.  

This paper is laid out as follows: In the following section (\ref{Preliminaries}) we will introduce a generic form for gravitation theories minimally coupled on a flat, Robertson-Walker geometry and establish some background results regarding symmetry. In section (\ref{LMeasure}) we introduce the Liouville measure in this context and show how its evolution acts on this symmetry. This brings us to some physical applications and the interpretation of relative phase space volume as a probability measure in section (\ref{PDFs}). Finally we conclude with some notes on application beyond minimal coupling. 

\section{Preliminaries} \label{Preliminaries}
Let us consider an action for gravity minimally coupled to matter on some Lorentzian manifold $M$:

\be S = \int_M \sqrt{g} F[R,R_{ab}] + \sqrt{g} \L_m [q,\dot{q}] \label{action} \ee

Here and throughout, the subscript $m$ shall refer to the matter, whose dynamics are determined by the matter Lagrangian, $\L_m$ through fields $q$. 

 In particular, let us
examine flat Robertson-Walker geometries which are coupled minimally to matter,
with the gravitational Lagrangian left largely unconstrained, being only a
function of the Hubble rate (the only gauge invariant geometrical quantity
available). Matter will be defined in terms of fields, coordinatized by $q_i$: 

\be \L = v (g[H]+\L_m [q_i,\dot{q_i}]) \label{MCL} \ee

For choices of the function $g[H]$ this can represent General Relativity ($g[H]=H^2$), $F(R)$ theory, effective Loop Quantum Cosmology \cite{Ashtekar:2008zu} etc. Indeed, since the Hubble rate is the only geometrical observable, any theory should be cast in the mold. The fields are minimally coupled as $v$, the volume of some fiducial cell, multiplies each term. Let us further demand that $g$ be differentiable with invertible derivative \footnote{This is not a highly restrictive demand in the space of theories, as these requirements are such needed to give a well defined action formulation in the first place.}. Then we immediately obtain:

\be b = g'[H] \:\:\:\: P_i = v \f{\partial \L_m}{\partial \dot{q_i}} =vp_i \:\:\:\: \dot{q_i} = \Phi_i[\f{P_j}{v}] \ee

In which the $\Phi_i$ determine uniquely the $\dot{q_i}$ or are constraints, and $b$ represents the momentum conjugate to volume. The symplectic structure is given:

\ba \omega &=& db \wedge dv + dp \wedge dq \nonumber  \\
		   &=& db \wedge dv + v dp_m \wedge dq + p_m dv \wedge dq \nonumber \\
		   &=& (db-p_m dq) \wedge dv + v \omega_m \label{symp} \ea

Hence we derive the Hamlitonian

\be \label{Hmincoup} \H = v(b g'[b]^{-1}-g[g'[b]^{-1}] + \f{P_i}{v} \Phi_i[\f{P_j}{v}] -\L_m[q_i, \Phi_i[\f{P_j}{v}]) \ee

Which, upon gathering terms is of the form:

\be \label{H} \H = v(A[b] + H_m[p_i,q_i]) \ee

wherein $p_i=P_i/v$. Note that there is a useful separation between the gravitational degree of freedom, encoded in $P_v$ and the matter which contributes to $H_m$. The notation has been chosen to indicate that one can simply make the substitution of momenta in the matter Hamiltonian. At this point we are simply considering the behaviour of a Robertson-Walker geometry, however, anisotropies can also be encoded within this when dealing with a Bianchi I cosmology \cite{Ashtekar:2009vc}. 

From (\ref{Hmincoup}) and (\ref{symp}) it is immediately apparent that there exists a symmetry of our system corresponding to rescaling the volume whilst keeping its conjugate momentum ($b$) and the matter coordinates ($q$) and matter momenta ($p_m$) fixed. This symmetry exists because on we have performed the spatial integrals in the action (\ref{action}) over a fiducial cell whose size is arbitrary - it is a fundamental symmetry of the system that choosing a different cell, and hence rescaling $v$ by a constant, should lead to identical dynamics for the observable degrees of freedom.

Define the vector field $G$ by

\be G = v \f{\partial}{\partial v} +  \sum_i P_i \f{\partial}{\partial P_i} \ee

Then we find that $G$ generates a symmetry between solutions. Under the action of this vector field, the matter degrees of freedom (and hence matter Hamiltonian and symplectic structure) are conserved: $\Lie_G H_m = 0 = \Lie_G \omega_m$.  $G$ commutes with the Hamiltonian flow, and its action on the constraint and symplectic structure are given $\Lie_G H = H$,$\Lie_G \omega = \omega$

\be \Lie_G \H = \H \:\:\:\: \Lie_G \omega = \omega \:\:\:\: \Lie_G X_\H = 0 \ee

where $X_H$ is the vector field generated by the Hamiltonian flow. 

As a further consequence solely of minimal coupling, the continuity equation is automatically satisfied:

\ba \dot{\rho} &=& \f{\partial \rho}{\partial P} \dot{P} + \f{\partial \rho}{\partial q} \dot{q} +\f{\partial \rho}{\partial v} \dot{v} \nonumber \\
	       &=& -\f{\partial \rho}{\partial P} \f{\partial \H}{\partial q} + \f{\partial \rho}{\partial q}\f{\partial \H}{\partial P} + \f{\dot{v}}{v}(\rho+P_r) \nonumber \\
	       &=& -v \f{\partial \rho}{\partial P} \f{\partial \rho}{\partial q} + v \f{\partial \rho}{\partial q}\f{\partial \rho}{\partial P} +3H(\rho+P_r) \nonumber \\
	       &=& 3H(\rho+P_r) \ea

Further the volume momentum is monotonic non-increasing when considering matter which obeys the weak energy condition:

\be \dot{b} = -\f{\partial \H}{\partial v} = - \f{\H}{v}- v \f{\partial \rho}{\partial v} = -(\rho+P) \ee

\section{The Liouville Measure} \label{LMeasure}

In order to perform any solution counting we require a measure on phase space. One such measure which is readily available with the tools already introduced is the Liouville measure. Before continuning with our analysis of minimally coupled systems, let us recount some properties of this measure.

Our phase space is a symplective manifold on which Liouville's theorem states that the volume of phase space occupied by a set of dynamical trajectories as measured by the Poincare invariant (the top power of the symplectic structure) is invariant under evolution, ie $\L_{X_H} \omega^n = 0$. In fact, the symplectic structure itself is conserved, not only its highest power.

The measure is invariant under coordinate transformations, a reparametrization of our system. Such reparametrizations are of great importance as, in the absence of external input, there is often no natural choice of parametrization of a physical system. If we are told simply that a free parameter in our system takes a value in some (interval of a) field, for example, a suitable choice of parametrization can transform between any pair of chosen priors for such a parameter \cite{Norton}. Our system is described by a Lagrangian $\L[x,\dot{x}]$. If we make a change of variable to $w[x]$ we find:

\be P_x = \f{\partial \L}{\partial \dot{x}} = \f{\partial \L}{\partial \dot{w}} \f{dw}{dx} = P_w w' \ee

Hence 

\be \omega = dw \wedge dP_w = w' dx \wedge d\f{P_x}{w'} = dx \wedge dP_x \ee

and we are lead to the same measure regardless of the choice of parametrization. 

To evaluate the number of solutions to our minimally coupled system, we must consider a surface which each solution crosses exactly once. Since we have established that $P_v$ is monotonic, setting surfaces of constant $P_v$ will perform this role. \footnote{In doing so we ignore solutions for which $P_v$ is a constant - i.e. those which are pure de-Sitter - these solutions exist when the matter content is purely a cosmological constant, and therefore are not of interest to general dynamical systems.} Our space of solutions is therefore:

\be \S_C = \H^{-1} [0] \cap b^{-1} [C] \ee

and Liouville's theorem assures us of the invariance of this measure under changes in the choice of $C$. On such surfaces, $\rho$ is constant and the symplectic structure can be expressed:

\be \overleftarrow{\omega} = \sum_i dq_i \wedge dP_i = \sum_i v dq_i \wedge dp_i + \sum_i p_i dq_i \wedge dv \ee

Thus, raising this to the $n$th power to form our measure. Note that since $\rho$ is a function of the $q_i$ and $p_i$ which is constant, then not all of the $dq_i$ and $dp_i$ can be orthogonal. In other words, we can use the constancy of matter energy density on this surface to determine one of the momenta in terms of the remaining phase space coordinates. Repeating this process term by term in the $2n$-form we find:

\be \overleftarrow{\omega^n} = \sum_i v^{n-1} p_i dv \wedge dq_i \wedge \prod_{j \neq i} (dq_j \wedge dp_j) \ee

In each term $p_i$ can be expressed in terms of the remaining phase-space coordinates. 

Let us now turn our attention to the topology of the space of solutions, $S_C$. In particular, we shall examine the effects of minimal coupling gravity when the space of solutions to the pure matter Hamlitonian at a fixed energy, $S_m(E)$ is compact. Although it is possible to perform analysis on the dynamical behaviour of solutions on a non-compact space, our ultimate aim will be to define fractions of phase-space volume on which solutions exhibit certain properties. In non-compact spaces there is a further ordering problem of how the counting of solutions is performed which will not be analysed here, and hence we shall confine ourselves to the relatively simpler scenario. Since there exists a scaling freedom in $v$, the total space of solutions is non-compact. However, the existence of the non-compact gauge direction allows us to remove this. 

Let us define the (non-canonical) coordinate $R$ on phase space by 

\be \label{R} R^2 = v^2 + \sum_i P_i^2 \ee

Thus our gauge direction is parametrized by $R$: 

\be \label{GR} G = v \f{\partial}{\partial v} + \sum_i P_i \f{\partial}{\partial P_i} = R \f{\partial}{\partial R} \ee

By further introducing angular coordinates $\theta, \psi_i$ we can let $v=R cos[\theta]$ and $P_i = R sin[\theta] T_i[\overrightarrow{\psi}]$. Here $T_i$ is a decomposition of the unit sphere in angular coordinates given:

\ba T_1 &=& cos[\psi_1] \\ 
    T_i &=& sin[\psi_1]...sin[\psi_{i-1}] cos[\psi_i] \\
    T_n &=& sin[\psi_1]...sin[\psi_{n-1}] \ea
    
Thus our Hamiltonian constraint can be written: 

\be \label{HR} H= R cos[\theta] (A[b] + H_m[q_i, tan[\theta]T_i[\overrightarrow{\psi}]]) \ee

Thus it is apparent that the topology of $S_C$ is inherited directly from $H_m$, being $S_m(A[b^{-1}[C]]) \times \mathbb{R}_+$. Thus if $S_m$ is a compact space, the space of physically distinct solutions of our system will be compact, with a real degree of freedom in the gauge direction. We could, at this stage, project our measure onto these coordinates and consider the gauge direction, as on the space of solutions $dv=dR/cos[\theta]$. Eventually to form a useful measure on will have to `project out' the gauge direction, usually be fixing an interval in $R$ or $v$ over which to perform an integral, and this choice will lead to distinct measures. This is normally be carried out by selecting $v$, however since $R$ represents the gauge direction more completely, this leads to a simplification of structures which will be highlighted below.

Phase space measures are typically employed to answer questions regarding inflationary cosmology \cite{Gibbons:1986xk,Gibbons:2006pa,Measure,Measure2,Kaya:2012xq,Remmen:2013eja,Wald}. Therefore, let us consider the canonical example of GR coupled to a massive scalar field which will perform the roll of inflaton. For brevity of expression we shall set all masses, physical constants etc to unity. Our matter Lagrangians are 

\be \label{Lex} L_m = \f{\phi^2}{2} + \f{\dot{\phi^2}}{2} \:\:\: L_g = - \f{\dot{v^2}}{2v} \ee 

and so $b =\f{\dot{v}}{v}$ and $P=v\dot{\phi}$. Hence in the original variables we find our Hamiltonian and measure

\be \label{Hex} H=v(-\f{b^2}{2} +\f{P^2}{2v^2} + \phi^2) \:\:\:\: \overleftarrow{\omega}=\sqrt{b^2-\phi^2}d\phi\wedge dv\ee

In terms of the gauge direction, we let $v=Rcos(\theta) \: P=Rsin(\theta)$ and these can be expressed:

\be \label{HexR} H=R cos(\theta) (-\f{b^2}{2} + tan^2[\theta] + \phi^2) \:\:\:\: \overleftarrow{\omega} = \f{\sqrt{b^2-\phi^2}}{\sqrt{1+b^2-\phi^2}} d\phi\wedge dR\ee

On a given portion of phase space, these two measures are identical; however when projecting out by taking an integral over a fixed range in $R$ as opposed to a fixed range in $v$ the resulting measures differ. This generalizes the results presented in \cite{Corichi:2013kua}.

Generalizing to $n+1$ (possibly interacting) scalar fields, our Hamiltonian is 

\ba \label{Hexm} H &=& v(-\f{b^2}{2} +\sum_i \f{P_i^2}{2v^2} +V[\overrightarrow{\phi}]) \\
				   &=& R cos[\theta] ( -\f{b^2}{2} +tan^2[\theta] + V[\overrightarrow{\phi}]) \ea

Here we see the role of $R$ and $\theta$ made more explicit - $tan[\theta]$ represents to split between kinetic and potential energy in the system at a fixed $b$, and the angular coordinates parametrize the distribution of kinetic energy across the differing field momenta. Since there is an $S_n$ symmetry in this choice, the individual $\psi$ do not appear in the Hamiltonian. These multifield models of inflation lead to a range of differing physical outcomes depending on the specific choice of potential and interactions \cite{Peterson:2010np,Easther:2013bga,Easther:2013rva} and exhibit attractors \cite{Kallosh:2013daa}.   

On examination of the symplectic structure we again see the symmetry in distribution of kinetic energy made explicit: 

\be \overleftarrow{\omega} = R^{n} sin^{n+1} [\theta] \sum_k \left( T_k dR \wedge d\phi_k \prod_{i \neq k} dT_i \wedge d\phi_i \right) \ee

To simplify this somewhat we note that the Hamiltonian constraint again allows us to rewrite $sin[\theta]$ in terms of $b$ and $V[\overrightarrow{\phi}]$. Furthermore, since each of the $T_i$ with $i<n$ contain a term proportional to $T_{i-1}$ multiplied by a term in $\psi_i$ we can expand the product of $dT_i$ and multiply out to obtain the unit measure on the n-sphere $S_n$ thus:

\be \overleftarrow{\omega} = R^{n} \left(\f{b^2 - V[\overrightarrow{\phi}]}{1+b^2 - V[\overrightarrow{\phi}]} \right)^{\f{n+1}{2}} dR \wedge dS_n \wedge d\overrightarrow{\phi} \ee

in which $dS_n$ is the metric on the unit n-sphere reflecting the angles $\psi$, and $d\overrightarrow{\phi}$ represents $d\phi_1 \wedge ... \wedge d\phi_{n+1}$. 

Thus for multi-field `n-flation' models there is a clear separation of the measure into distributing the matter energy density between kinetic and potential. Furthermore, for theories of gravity other than pure GR, the direct substitution of $b^2 \rightarrow E[b]$, where $E=A^{-1}[b]$ is the relevant correction to the Friedmann relation between energy and gravitational momentum as in eq. \ref{H}. In particular, if we wish to evaluate the section of phase-space, after gauge-fixing $R$, on which the fraction of matter energy which is potential is greater than $Q$, say, we find:

\be P(PE < Q)= \f{\int_{A[V]} d\overrightarrow{\phi} \left(\f{E-V[\overrightarrow{\phi}]}{1+E-V[\overrightarrow{\phi}]}\right)} {\int_{C[V]} d\overrightarrow{\phi} \left(\f{E-V[\overrightarrow{\phi}]}{1+E-V[\overrightarrow{\phi}]}\right)} \ee

in which $A[V]$ represents that fraction of the potential on which the condition is met, and $C[V]$ is the total area in which the potential energy is below the total energy. Thus this condition depends solely on the shape of the potential $V$ and is independent of the particular choice of gravitational theory. 

\section{Induced Probability Density Functions} \label{PDFs}

Let $\mathbb{M}$ be a manifold, which can be decomposed into as the product of two separate manifolds: 

\be \mathbb{M}=\mathbb{S} \times \mathbb{G} \ee

In which $\mathbb{S}$ is compact, and $\mathbb{G}$ is non-compact. Further, let $\Xi_M$ be a volume form on $\mathbb{M}$: From this one can induce a normalized volume form $\Xi_S$ on $\mathbb{S}$ by

\be \Xi_S = \f{\int_\mathbb{G} \Xi_M}{\int_\mathbb{M} \Xi_M} \ee

However, since $\mathbb{G}$ is non-compact, any integral is calculated as the limit of integrals over compact subspaces $\mathbb{G}=\cup G_i$. Freedom of union and repartition of these subspaces means that any separate sequence $G^n_i$ is equally valid, and therefore inequivalent volume forms can be created by using differing choices of partitioning $\mathbb{G}$ at separate points on $\mathbb{S}$. Let $\mathbb{T}$ be the space of such sequences. One can therefore create a family of inequivalent measures 

\be \Xi_S^n (x) = \lim_{i \to \infty} \f{\int_{g_i} \Xi_M}{\int_\mathbb{M} \Xi_M} \ee

In which $g_i : \mathbb{S} \rightarrow{\mathbb{T}}$ is a choice of sequence for each point in $\mathbb{S}$. Thus the projection of a measure on a non-compact space onto a compact subspace is not unique. As an example let us consider $\mathbb{S}=[0,1]$ and $\mathbb{G}=\mathbb{R}$. Then the measure $dxdy$ on $M$ seems a natural choice for $\Xi_M$ and if we let $G_i$ be independent of choice of $x$ we recover through this process the usual measure $dx$ on $[0,1]$. However, we could equally have chosen to cover $\mathbb{R}$ by choosing:

\be g_\lambda(x) = [-\lambda f(x),\lambda f(x)] \ee

For any always positive function $f$, and taking the limit $\lambda \rightarrow \infty$ we recover the measure 

\be \Xi'=\f{f(x)dx}{\int_0^1 f(x)dx} \ee

Thus it is clear that by choice of limiting procedure in the non-compact direction, one can induce any volume form on a compact submanifold. 

\subsection{Induced Physical Measures} \label{PMs}

In considering the Liouville measure in the case of minimally coupled fields, recall that we do indeed have a non-compact gauge direction, $R$. If we wish to induce a measure on only the physically relevant variables we must integrate out this freedom. However, as was shown above, integrating over this non-compact direction can induce differing measures depending on how limits are taken. The key motivation behind the use of the Liouville measure is its preservation under evolution, however although the volume of any section of phase space is conserved, the area obtained by projecting onto a subset of variables is not, unless the gauge direction evolves uniformly \cite{Corichi:2010zp}. 

Let $r=R^{n+1}$. Then $\overleftarrow{\omega} = f(K) dr \wedge dK$ where $K$ comprise the physically relevant variables comprising the remaining degrees of freedom in our system. We would like to project out the gauge direction, to form a measure $\chi = f(K) dK$. Liouville's theorem states that $\Lie_{X_H} \omega =0$ where $X_H$ is the flow induced by the Hamiltonian. Indeed it would be sufficient for our purposes, since we normalize, if $\Lie_{X_H} \chi \sim \chi$. However, since $r$ will evolve we find: 

\ba 0 &=& \Lie_{X_H} \overleftarrow{\omega} \nonumber \\
      &=& dr \wedge \Lie_{X_H} \chi + \Lie_{X_H} (dr) \wedge \chi \nonumber \\
      -dr \wedge \Lie_{X_H} \chi &=& d(\Lie_{X_H} r) \wedge \chi \label{MEvolve} \ea

Therefore this procedure is unique only when $\Lie_{X_H} r$ is independent of initial choices of variables $K$. In general this is not true: 

\ba \Lie_{X_H} r &=& \{H,r\}  \nonumber \\
                 &=& cos[\theta] \f{\partial H}{\partial b} + sin[\theta] \f{\partial H}{\partial q_i} T_i \ea

This can contain terms proportional to any of the physically relevant variables, and therefore one cannot unambiguously project the Liouville measure. Since Liouville's theorem holds, we can use this to explain the existence of attractors in phase space. Consider forming a measure by projecting out the gauge direction by integrating over a fixed interval, $I$ at some initial $b$. Since the volume of phase space as measured by $\overleftarrow{\omega}$ is conserved throughout the evolution, any change in the length of $I$ is compensated by a change in the distribution over physically relevant variables. In particular, by performing this integration on \ref{MEvolve} we find:

\ba \Lie_{X_H} \chi &=& \int_I d\Lie_{X_H}r \chi \nonumber \\
                    &=& \int_I dU(k) \chi \nonumber \\
                    &=& p(K) \chi \ea

 in which $U=\Lie_{H_H}r$. Therefore between two points, $b=b_i$ and $b=b_f$ say, $\chi$ is not conserved. Thus we find that the expansion in the gauge direction can be expressed as a change in measure, in effect bringing about a probability density function $P(K)=\int_{b_i}^{b_f} p(K) db$, whose magnitude is determined, up to overall normalization, by the expansion of the extent of the gauge direction. In particular we find that $P(K)$ is largest on those solutions which expand the most in the gauge direction, and thus our measure becomes most focused on those solutions. Here, if the potential is unbounded above \footnote{If the potential is bounded from above, the situation becomes more subtle: Maximizing expansion in the gauge direction over all configurations may not allow access to complete potential domination ($\theta=0$) and thus focusing may occur on those solutions which expand the gauge direction most due to gradients in the potential. The general procedure is to seek to maximize $\f{dR}{db}$ across solutions between start and end points.} those solutions can be determined to be those on which $\theta$ is minimized - i.e. solutions which have the highest potential energy, and hence the greatest expansion. We therefore recover a version of ``volume weighting'' \cite{Linde:2007nm,Hawking:2007vf,Winitzki:2008yb} of solutions - those solutions which undergo the greatest expansion are dynamically attractors . 

\section{Conclusions} \label{Conclusions}

The existence of a gauge symmetry in the Hamiltonian formulation of cosmology is apparent, regardless of the particular theory in question. This symmetry is basis of the attractor behaviour apparent in dynamical systems. The existence of attractors, although it would initially appear to contradict Liouville's theorem, in fact is a direct consequence of said theorem. It is the manner in which gauge directions are projected out that leads to this phenomenon. 

Furthermore, the solutions which appear as attractors are those which undergo the greatest expansion. This is a theory agnostic result, arising from nothing more than the principle of minimally coupling matter to a gravitational action. The expansion of the space of solutions along the gauge direction is compensated by the convergence of solutions on these attractors. In theories which approximate GR at low energies, such solutions will be those which are almost de-Sitter. 

Although here we have focussed on minimally coupled systems, many of the results extend naturally to non-minimal coupling. The significant role played by minimal coupling is that the conjugate momentum to volume is, given that matter obeys energy conditions, monotonic. Thus we can treat this as a clock, and use its constancy to provide a surface on which to count solutions. Furthermore, due to the relationship between this variable and matter energy density this grants us direct access to physical observables on each trajectory, and thus we are able to ask questions of the distributions of observables at an identifiable event (e.g. a given value of the Hubble parameter). In the case of non-minimal coupling the distinction between matter and geometrical parameters becomes less clear, and without a specific theory one cannot tell if certain parameters would form a good clock. Thus one would be unable to know a priori if two sets of observations came from distinct trajectories. One way around this problem is to follow the methods of quantum cosmology \cite{Ashtekar:2006wn} and introduce a minimally coupled massless scalar field to act as a clock, whose momentum would be a constant of the motion. Thus observables could be evaluated on slices of constant time, as defined by this clock. However, in the absence of direct observations of the clock the choice of initial time at any configuration of physical parameters is arbitrary, and thus we once again encounter a problem of counting. 

Let us finish with some speculative remarks. Consider fields interacting subject to a potential $V(\phi)$, whose initial configuration is determined by some process at a high energy-density. As this density drops, these solutions are focused on those with the highest expansion rate. Local minima of $V$ will appear as cosmological constants for fields defined about such minima. Therefore one should expect attractors to be the highest cosmological constant available within the potential. Given a randomly selected potential, this would seem to greatly exceed the observed value. However, the observed value of the cosmological constant is close to the anthropic bound, and thus volume weighted attractors provide the natural counterpart to this: Anthropic considerations place an upper bound on $\Lambda$, and the attractor behaviour pushes solutions to this bound. 

\section*{Acknowledgements}

The author is indebted to comments from Josh Schiffrin. Support was provided by a grant from the Templeton Foundation. 
\bibliographystyle{unsrt}
\bibliography{InflationNotes}

\begin{thebibliography}{10}

\bibitem{Measure}
Abhay Ashtekar and David Sloan.
\newblock {Loop quantum cosmology and slow roll inflation}.
\newblock {\em Phys.Lett.}, B694:108--112, 2010.

\bibitem{Measure2}
Abhay Ashtekar and David Sloan.
\newblock Probability of inflation in loop quantum cosmology.
\newblock {\em General Relativity and Gravitation}, 43(12), 2011.

\bibitem{Corichi:2013kua}
Alejandro Corichi and David Sloan.
\newblock {Inflationary Attractors and their Measures}.
\newblock {\em Class.Quant.Grav.}, 31:062001, 2014.

\bibitem{Ashtekar:2008zu}
Abhay Ashtekar.
\newblock {Loop Quantum Cosmology: An Overview}.
\newblock {\em Gen.Rel.Grav.}, 41:707--741, 2009.

\bibitem{Ashtekar:2009vc}
Abhay Ashtekar and Edward Wilson-Ewing.
\newblock {Loop quantum cosmology of Bianchi I models}.
\newblock {\em Phys.Rev.}, D79:083535, 2009.

\bibitem{Norton}
John~D. Norton.
\newblock Cosmic confusions: Not supporting versus supporting not.
\newblock {\em Philosophy of Science}, 77(4):501--523, 2010.

\bibitem{Gibbons:1986xk}
G.W. Gibbons, S.W. Hawking, and J.M. Stewart.
\newblock {A Natural Measure on the Set of All Universes}.
\newblock {\em Nucl.Phys.}, B281:736, 1987.

\bibitem{Gibbons:2006pa}
G.W. Gibbons and Neil Turok.
\newblock {The Measure Problem in Cosmology}.
\newblock {\em Phys.Rev.}, D77:063516, 2008.

\bibitem{Kaya:2012xq}
Ali Kaya.
\newblock {Comments on the Canonical Measure in Cosmology}.
\newblock {\em Phys.Lett.}, B713:1--5, 2012.

\bibitem{Remmen:2013eja}
Grant~N. Remmen and Sean~M. Carroll.
\newblock {Attractor Solutions in Scalar-Field Cosmology}.
\newblock 2013.

\bibitem{Wald}
Joshua~S. Schiffrin and Robert~M. Wald.
\newblock {Measure and Probability in Cosmology}.
\newblock {\em Phys.Rev.}, D86:023521, 2012.

\bibitem{Peterson:2010np}
Courtney~M. Peterson and Max Tegmark.
\newblock {Testing Two-Field Inflation}.
\newblock {\em Phys.Rev.}, D83:023522, 2011.

\bibitem{Easther:2013bga}
Richard Easther and Layne~C. Price.
\newblock {Initial conditions and sampling for multifield inflation}.
\newblock {\em JCAP}, 1307:027, 2013.

\bibitem{Easther:2013rva}
Richard Easther, Jonathan Frazer, Hiranya~V. Peiris, and Layne~C. Price.
\newblock {Simple predictions from multifield inflationary models}.
\newblock {\em Phys.Rev.Lett.}, 112:161302, 2014.

\bibitem{Kallosh:2013daa}
Renata Kallosh and Andrei Linde.
\newblock {Multi-field Conformal Cosmological Attractors}.
\newblock {\em JCAP}, 1312:006, 2013.

\bibitem{Corichi:2010zp}
Alejandro Corichi and Asieh Karami.
\newblock {On the measure problem in slow roll inflation and loop quantum
  cosmology}.
\newblock {\em Phys.Rev.}, D83:104006, 2011.

\bibitem{Linde:2007nm}
Andrei~D. Linde.
\newblock {Towards a gauge invariant volume-weighted probability measure for
  eternal inflation}.
\newblock {\em JCAP}, 0706:017, 2007.

\bibitem{Hawking:2007vf}
S.W. Hawking.
\newblock {Volume Weighting in the No Boundary Proposal}.
\newblock 2007.

\bibitem{Winitzki:2008yb}
Sergei Winitzki.
\newblock {A Volume-weighted measure for eternal inflation}.
\newblock {\em Phys.Rev.}, D78:043501, 2008.

\bibitem{Ashtekar:2006wn}
Abhay Ashtekar, Tomasz Pawlowski, and Parampreet Singh.
\newblock {Quantum Nature of the Big Bang: Improved dynamics}.
\newblock {\em Phys.Rev.}, D74:084003, 2006.

\end{thebibliography}

\end{document}